\documentclass{article}
\usepackage{cite}
\usepackage{amsmath,amssymb,amsfonts}
\usepackage{algorithmic}
\usepackage{graphicx}
\usepackage[width=18cm,height=25cm]{geometry}
\usepackage{textcomp}
\usepackage{siunitx}
\usepackage{tikz}
\usepackage{pgfplots}
\pgfplotsset{width=9cm,compat=1.9}
\usepackage{eso-pic}

\AddToShipoutPicture*{\footnotesize\sffamily\raisebox{1cm}{\hspace{1.65cm}\fbox{\parbox{\textwidth}{\copyright 2023 IEEE.  Personal use of this material is permitted.  Permission from IEEE must be obtained for all other uses, in any current or future media, including reprinting/republishing this material for advertising or promotional purposes, creating new collective works, for resale or redistribution to servers or lists, or reuse of any copyrighted component of this work in other works.}}}}

\title{Data-Driven Update of B(H) Curves of Iron Yokes 
in Normal Conducting Accelerator Magnets}
\author{Luisa Fleig, Melvin Liebsch, Stephan Russenschuck,\\ Sebastian Schöps}
\date{June 2023}

\begin{document}

\maketitle

\begin{abstract}
Constitutive equations are used in electromagnetic field simulations to model a material response to applied fields or forces. The $B(H)$ characteristic of iron laminations depends on thermal and mechanical stresses that may have occurred during the manufacturing process. Data-driven modelling and updating of the $B(H)$ characteristic are therefore well known necessities. 
In this work the $B(H)$ curve of an iron yoke of an accelerator magnet is updated based on observed magnetic flux density data by solving a non-linear inverse problem. The inverse problem is regularized by restricting the solution to the function space that is spanned by the truncated Karhunen Loeve expansion of a stochastic $B(H)$-curve model based on material measurements. 
It is shown that this method is able to retrieve a previously selected ground truth $B(H)$-curve. With the update of the $B(H)$ characteristic, the numerical model gains predictive capacities for excitation currents that were not included in the data.
\end{abstract}


\section{Introduction}
\label{sec:introduction}
Numerical field computation is an established tool for the design of accelerator magnets and for predicting the field for a given excitation current.
The numerical models depend on constitutive parameters, such as the $B(H)$ curve of the yoke material, neglecting hysteresis.
For material specimens, a discrete evaluation table of this curve can be measured with a split coil permeameter \cite{Arpaia}. 
Approaches to determine a continuous and monotone $B(H)$ curve from the discrete evaluation table using splines can be found in \cite{Reitzinger} and \cite{Pechstein}. An alternative, spline-based method using magnetic flux measurements is described in \cite{Kaltenbacher}.
In addition, there are various closed-form expressions such as the Wlodarski curve \cite{Wlodarski}, whose parameters can be fitted to measurements \cite{Sorti, Abou, Mohammadi}. Experience has shown that fitting these closed form expressions to the measured permeameter data results in large residuals of up to 10\%. 
\newline Due to variations in mechanical stresses \cite{Mohammadi} applied  to the material specimens in the manufacturing process, as well as ageing effects, the $B(H)$ curve varies both within a set of specimens of the same material and between specimens and the magnet yoke itself \cite{Cranganu}. This observation results in two approaches. First, measurement data of electromagnetic devices as built are used to update the $B(H)$ curve in the numerical model \cite{Sorti,Abou,Mohammadi}. Second, alternative models of the $B(H)$ dependence, such as point clouds \cite{deGersem}, or data-driven stochastic models \cite{RoemerPaper} are used. Both approaches are based on first and second principle (data-driven) modelling. This is in the spirit of hybrid modelling \cite{Kurz}.
\newline In this paper the data-driven update of the $B(H)$ curve of a magnet yoke is based on the stochastic $B(H)$ curve model of \cite{RoemerPaper}, which best describes the measured $B(H)$ curve fluctuations. The truncated Karhunen Loeve expansion used in the stochastic $B(H)$ curve model regularizes the inverse problem of retrieving the $B(H)$ curve from observed magnetic flux density data.
\newline To select optimal positions in the magnet to collect magnetic flux density data, the sensitivity analysis described in \cite{Roemer} is applied. Subsequently, it is shown that the proposed method has the capacity to retrieve a previously selected ground truth $B(H)$ curve that was used to generate the magnetic flux density data. The resulting updated numerical model, tailored to a specific magnet as built, allows interpolative field prediction for excitation currents not considered in the update.

\section{Methods}
We consider a normal-conducting iron dominated magnet, with a non-laminated yoke of homogeneous material. Anisotropy, hysteresis, dynamic and coupled effects are neglected in the magnet model.

\subsection{The magnetostatic field problem}

The numerical model of the normal-conducting, iron dominated magnet on a domain $D$ is based on the magnetostatic problem given by
$\mathrm{curl}\: \mathbf{H}=\mathbf{J}$ in $D$, $\mathrm{div}\:\mathbf{B}=0$ in $D$ and $\mathbf{B}\cdot \mathbf{n}=0$ on $\partial D$. Thereby, $\mathbf{B}$ is the magnetic flux density, $\mathbf{H}$ denotes the magnetic field strength, $\mathbf{J}$ is the electric current density with $\mathrm{div}\:\mathbf{J}=0$ and $\mathbf{n}$ is the outward pointing unit normal vector. We assume that the domain $D$ is contractible. 

The magnetic vector potential $\mathbf{A}$ can be introduced such that $\mathrm{curl}\:\mathbf{A}=\mathbf{B}$. By incorporating the vector potential and the constitutive equation $\nu(\left\|\mathbf{B} \right\| ) \mathbf{B} = \mathbf{H}$, the curl-curl formulation of the magnetostatic problem
\begin{align}
\mathrm{curl}\:\nu(\left\| \mathrm{curl}\:\mathbf{A}\right\| )\mathrm{curl}\:\mathbf{A}&=\mathbf{J} & \mathrm{in}\; D\\
\mathbf{A}\times \mathbf{n}&=0& \mathrm{on}\; \partial D.\label{eq:curlcurl}
\end{align}
is obtained. The corresponding weak formulation is given by: Find $\mathbf{A}\in\mathcal{V}$, where
\begin{equation}
\mathcal{V}:=\left\lbrace \mathbf{v}\in H_0(\mathrm{curl};D)\: \middle|\: \left< \mathbf{v} , \mathrm{grad}\:w \right>_D = 0 \, \forall w\in H_0^1(D) \right\rbrace  
\end{equation}
such that 
\begin{equation}
\int_D\nu(\left\| \mathrm{curl}\:\mathbf{A}\right\| )\mathrm{curl}\:\mathbf{A}\cdot \mathrm{curl} \:\mathbf{v} \: \mathrm{d}V = \int_D \mathbf{J}\cdot \mathbf{v}\:\mathrm{d}V  \quad \forall \mathbf{v}\in \mathcal{V}. \label{eq:weak}
\end{equation}
A proof for the existence of a unique solution can be found in \cite{RoemerPaper}. In this work, the magnetostatic problem is discretized and solved in the FEM software getDP \cite{Geuzaine} in a two dimensional setting.

\subsection{Material model}
\label{sec:material}

In non-linear, isotropic and anhysteretic material the reluctivity function $\nu:\mathbb{R}^+_0\rightarrow\mathbb{R}^+_0$ is for $\mathbf{B}\neq \mathbf{0}$ defined by
\begin{equation}
\nu(\left\| \mathbf{B}\right\| ) =\frac{ f_\mathrm{HB}(\left\| \mathbf{B} \right\| )} {\left\| \mathbf{B} \right\|}.
\end{equation}
The material curve $f_\mathrm{HB}:\mathbb{R}^+_0\rightarrow\mathbb{R}^+_0, \: B \mapsto H $ is the bijective monotone function that maps the intensity $B:=\left\| \mathbf{B}\right\|$ of the magnetic flux density to the intensity $H:=\left\| \mathbf{H}\right\|$ of the corresponding magnetic field strength. Further physical properties, that are also essential for the existence of a unique solution of the corresponding weak magnetostatic problem, can be found in \cite{PechsteinThesis}. The inverse $f_\mathrm{HB}^{-1}$ is also known as $B(H)$ curve.
\newline If a set of $K$ material specimens is available, a split coil permeameter \cite{Arpaia} can be used to evaluate the function $f^k_\mathrm{HB}$ of each specimen on a discrete value set $0=B_1<\dots<B_L$. Due to the monotonicity of $f^k_\mathrm{HB}$
\begin{equation}
0=f^k_\mathrm{HB}(B_1)<\dots<f^k_\mathrm{HB}(B_L) \quad \forall 1\leq k \leq K.
\end{equation}
From these discrete material measurements, $K$ monotone functions 
\begin{equation}
f_k:I:=[0,B_L] \rightarrow \mathbb{R}^+_0 \quad k=1,\dots,K
\end{equation}
are derived using monotone cubic splines \cite{Fritsch}. 

Following \cite{RoemerPaper}, a probability space $(\Omega,\mathcal{F},\mathbb{P})$ is introduced that reflects the manufacturing-related variations of the material curve $f_{\mathrm{HB}}$. By abuse of notation, $f_\mathrm{HB}$ is a random field
\begin{equation}
f_\mathrm{HB}:\Omega\times I\rightarrow\mathbb{R}^+_0.
\end{equation}
According to \cite{RoemerPaper}, the weak magnetostatic problem is still uniquely solvable if the properties reported in \cite{PechsteinThesis} hold almost everywhere and if there is an $\omega$-independent lower bound $\alpha>0$ such that $f_\mathrm{HB}'(\omega,s)>\alpha$ for all $s\in I$ and almost all $\omega\in\Omega$. The functions $f_k$ derived from the material measurements are realizations of the random field in this setting. 
%karhunen loeve 
\newline Following \cite{RoemerPaper}, the random field $f_\mathrm{HB}$ is discretized with the truncated Karhunen Loeve expansion (KLE)
\begin{equation}
f^M_\mathrm{HB}(\omega,s)=\mathbb{E}[f_\mathrm{HB}(s)]+\sum_{m=1}^{M} \sqrt{\lambda_m} \:Y_m(\omega) \: b_m(s), \label{eq:KLE}
\end{equation}
where $(\lambda_m,b_m)$ with $\lambda_1>\dots>\lambda_M$ are eigenpairs of the operator $T_{f_\mathrm{HB}}:L^2(I)\rightarrow L^2(I)$ defined by
\begin{equation}
T_{f_\mathrm{HB}}(u)(s):=\int_I \mathrm{Cov}(f_\mathrm{HB}(s),f_\mathrm{HB}(t)) u(t) \:\mathrm{d}t
\end{equation}
and $Y_m$ are random variables that are for $\lambda_m>0$ given by
\begin{equation}
Y_m(\omega)=\frac{1}{\sqrt{\lambda_m}}\int_I\left( f_\mathrm{HB}(\omega,s)-\mathbb{E}[f_\mathrm{HB}(s)]\right)  b_m(s)\: \mathrm{d}s.
\label{Y_n}
\end{equation}
The eigenpairs $(\lambda_m,b_m)$ are obtained by solving the eigenvalue problem $T_{f_\mathrm{HB}}b=\lambda b$ in its weak form
\begin{equation}
\int_I\int_I  \mathrm{Cov}(f_\mathrm{HB}(s),f_\mathrm{HB}(t))b(s) v(t) \:\mathrm{d}s\mathrm{d}t = \int_I \lambda b(s) v(s) \mathrm{d}s
\end{equation}
using the Galerkin method and an approximation with radial basis functions. The eigenfunctions $b_m$ are also referred to as \textit{modes}.

Depending on the absolute value of the eigenvalues $\lambda_m$ the truncation threshold $M$ of the truncated KLE can be chosen such that truncation error is negligible and $f^M_\mathrm{HB}$ fulfills the essential properties for the existence of an unique solution of the weak magnetostatic problem.
Interpreting the random variables $Y_m$ as real-valued parameters of the material curve, which is described by the truncated KLE, yields with abuse of notation the \textit{parameterized material model} $f_\mathrm{HB}:\mathbb{R}^M\times I\rightarrow \mathbb{R}^+_0$ defined by
\begin{equation}
f_\mathrm{HB}(\mathbf{y},s):=\hat{f}_\mathrm{HB}(s)+\sum_{m=1}^{M} \sqrt{\lambda_m} \:\mathbf{y}_m \: b_m(s).
\label{eq:model}
\end{equation}
By definition, this material model describes best the observed variances. The expected value $\mathbb{E}[f_\mathrm{HB}]$ is approximated by the sample mean $\hat{f}_\mathrm{HB}$ of the realizations $f_k$.

Realizations $Y^k_m$ of the random variables $Y_m$ can be obtained by inserting $f_k$ and $\hat{f}_\mathrm{HB}$ in \eqref{Y_n}. Thus, the distribution of the random variables $Y_m$ and reasonable lower and upper bounds $\mathbf{y}^\mathrm{min}, \mathbf{y}^\mathrm{max}\in\mathbb{R}^M$ of the parameters $\mathbf{y}$ are estimated by
\begin{equation}
\mathbf{y}^\mathrm{min}_m:=\min_{1\leq k\leq K} Y^k_m \qquad \mathrm{and}\qquad \mathbf{y}^\mathrm{max}_m:=\max_{1\leq k\leq K} Y^k_m.
\end{equation}
To ensure the required monotonicity of the resulting parameterized material curves $s\mapsto f_\mathrm{HB}(\mathbf{y},s)$, the monotonicity of the curve in each combination of upper and lower bounds of $\mathbf{y}_m$ is checked and the bounds are adjusted accordingly. Due to the intermediate value theorem, the monotonicity then holds for all $\mathbf{y} \in \mathcal{Y}:=[\mathbf{y}^\mathrm{min}_1,\mathbf{y}^\mathrm{max}_1]\times\ldots\times[\mathbf{y}^\mathrm{min}_M,\mathbf{y}^\mathrm{max}_M]$.

\subsection{Sensitivity analysis}
In \cite{Roemer}, the sensitivity of the magnetic vector potential $\mathbf{A}$ and quantities of interest $Q$ with respect to perturbations of the reluctivity is discussed. To this end, the Gateaux derivative $\mathbf{A}'$ of the mapping $\tilde{\nu}\mapsto \mathbf{A}[\nu+\tilde{\nu}]$ is studied, where $\nu$ is the nominal reluctivity in $D_{\mathrm{iron}}$ and $\tilde{\nu}$ is a small perturbation. If $\mathbf{A}$ solves \eqref{eq:weak}, the Gateaux derivative $\mathbf{A}'\in\mathcal{V}$ is the weak solution of
\begin{align}
&&-&\int_{D_{\mathrm{iron}}}\boldsymbol\nu_\mathrm{d}(\mathrm{curl}\:\mathbf{A})\mathrm{curl}\:\mathbf{A}'\cdot \mathrm{curl}\:\mathbf{v}\:\mathrm{d}V
\\&&-&\int_{D_{\mathrm{air}}}\nu_0\:\mathrm{curl}\:\mathbf{A}'\cdot \mathrm{curl}\:\mathbf{v}\:\mathrm{d}V
\\
&=&&
\int_{D_{\mathrm{iron}}}\tilde{\nu}(|\mathrm{curl}\:\mathbf{A}|)\mathrm{curl}\:\mathbf{A}\cdot \mathrm{curl}\:\mathbf{v}\:\mathrm{d}V  \: \forall \mathbf{v}\in\mathcal{V},
\end{align}
where  $\boldsymbol\nu_\mathrm{d}(\mathbf{r}):=\nu(|\mathbf{r}|)\mathbb{I}+\nu'(|\mathbf{r}|)/|\mathbf{r}|\cdot \mathbf{r}\otimes\mathbf{r}$ is the differential reluctivity tensor \cite{Roemer}. The Gateaux derivative $Q'$ of a quantity of interest $Q$ is subsequently obtained by inserting $\mathbf{A}'$ in the observation function corresponding to the quantity of interest.
\newline With this method at hand, the sensitivity of quantities of interest such as components of the magnetic flux density or multipoles with respect to the weighted modes $\sqrt{\lambda_m} b_m$ can be examined.

\subsection{Identification of parameters}
\label{sec:identify_param}

The determination of the parameters $\mathbf{y} \in \mathbb{R}^M$ of the parameterized material model $f_\mathrm{HB}(\mathbf{y},\cdot)$ is a non-linear inverse problem. This can be solved by optimization:
\begin{align}
	&\min_{\textbf{y}\in\mathcal{Y}}
 \quad
 g(\textbf{y}):=\sum_{p,n}
 \left\| \mathbf{B}_n^{\mathrm{data}}(\mathbf{x}_p)-\mathrm{curl}\:\mathbf{A}_{\mathbf{y},n}(\mathbf{x}_p)\right\| ^2 \label{eq:objective} \\
	&\textrm{s.t.: } \forall \mathbf{v}\in\mathcal{V} \nonumber\\
 &\int_{D}\nu(\left\| \mathrm{curl}\:\mathbf{A}_{\mathbf{y},n}\right\| )\mathrm{curl}\:\mathbf{A}_{\mathbf{y},n}\cdot \mathrm{curl}\:\mathbf{v} \:\mathrm{d}V
		=\int_{D}\mathbf{v}\cdot \mathbf{J}_n\:\mathrm{d}V\nonumber\\
	& \nu(s)=\begin{cases}
    1/s\left( \hat{f}_\mathrm{HB}(s)+\sum_{m=1}^{M} \sqrt{\lambda_m} \:\mathbf{y}_m \: b_m(s)\right)  &\mathrm{in}\;D_{\mathrm{iron}} \\ 
    1/ \mu_0 &\mathrm{in}\;D_{\mathrm{air}}\end{cases}\nonumber
\end{align}
where the vector potential $\textbf{A}_{\textbf{y},n}$ depends (implicitly) on the parameters $\textbf{y}$ and the $n$-th current density $\mathbf{J}_n$. The objective function samples positions $(\mathbf{x}_p)$ in the air gap of the magnet and current levels $I_n$ at which the magnetic flux density is sensitive to perturbations, and measured data $\mathbf{B}^{\mathrm{data}}_n(\mathbf{x}_p)$ is available. 
The problem is solved by swarm optimization. In each optimization step, the corresponding weak magnetostatic problem is solved in getDP \cite{Geuzaine}.

In our test cases the solvability of the inverse problem by reducing the parameter space by means of the truncated KLE was sufficient. If an explicit regularisation is required, for example a Tikhonov regularization \cite{Vogel} can be applied. It is obtained by adding a penalty to the objective function \eqref{eq:objective}
\begin{equation}
    g_\mathrm{reg}(\textbf{y}):= g(\textbf{y}) + a\left\| 
  \mathbf{y}-\mathbb{E}[Y]\right\|_{\mathrm{Cov}(Y)} ^2,
\end{equation}
where $a\in \mathbb{R}$ is a hyper parameter.

\section{Application and Results}

\begin{figure}
	\centering
	\input{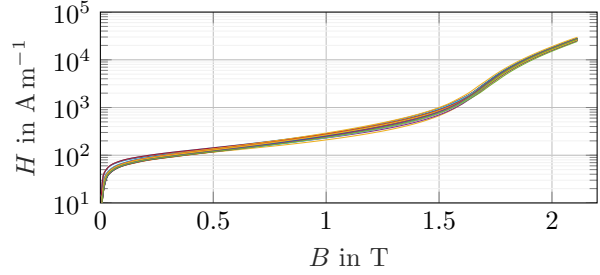}
	\caption{Monotone cubic splines $f_1,\dots,f_{26}$ obtained from ARMCO\textsuperscript{\textregistered} specimen measurements with a split coil permeameter.}
	\label{fig:matcurves}
\end{figure}

\begin{figure}
	\centering
	\input{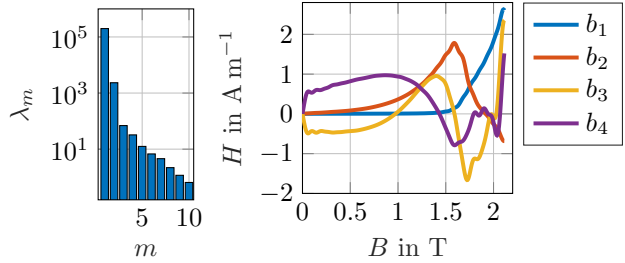}
	\caption{Ten largest eigenvalues $\lambda_m$ and eigenfunctions $b_1,\dots b_4$.}
	\label{fig:eigenpairs}
\end{figure}

With a split coil permeameter \cite{Arpaia} the $B(H)$-curves of $K=26$ toroidal ARMCO\textsuperscript{\textregistered} (American Rolling Milling Company) Pure Iron \cite{ARMCO} material specimens have been evaluated in $L=28$ points \cite{MarianoPhD}. 
Subsequently, the corresponding monotone functions $f_1,\dots,f_{26}$ (Fig.~\ref{fig:matcurves}) and eigenpairs $(\lambda_m,b_m)_{m=1,\dots,4}$ (Fig.~\ref{fig:eigenpairs}) are determined.
Due to the fast decay of the eigenvalues, the truncation threshold $M=4$ is chosen in the truncated KLE.

\begin{figure} 
\centering
		\def\svgwidth{0.5\linewidth}
		\small
		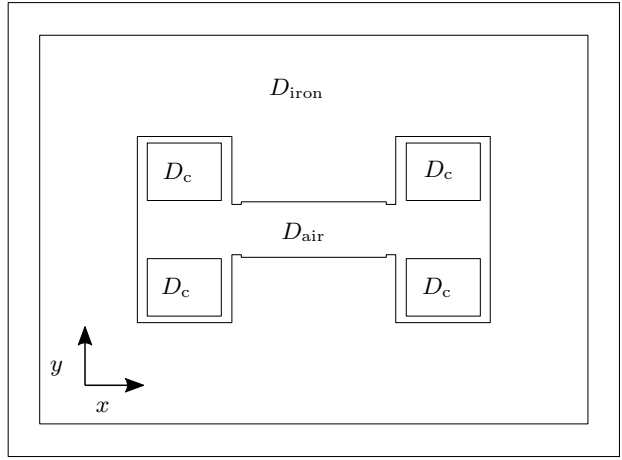  
			 \caption{Domain $D$ with the yoke $D_{\mathrm{iron}}$ of a H-shaped dipole and conductors $D_{\mathrm{c}}$ with $\mathrm{supp}(\mathbf{J})=D_{\mathrm{c}}$.\label{fig:yoke}}   
\end{figure}

\begin{figure} 
\centering
		\def\svgwidth{0.5\linewidth}
		\scriptsize
        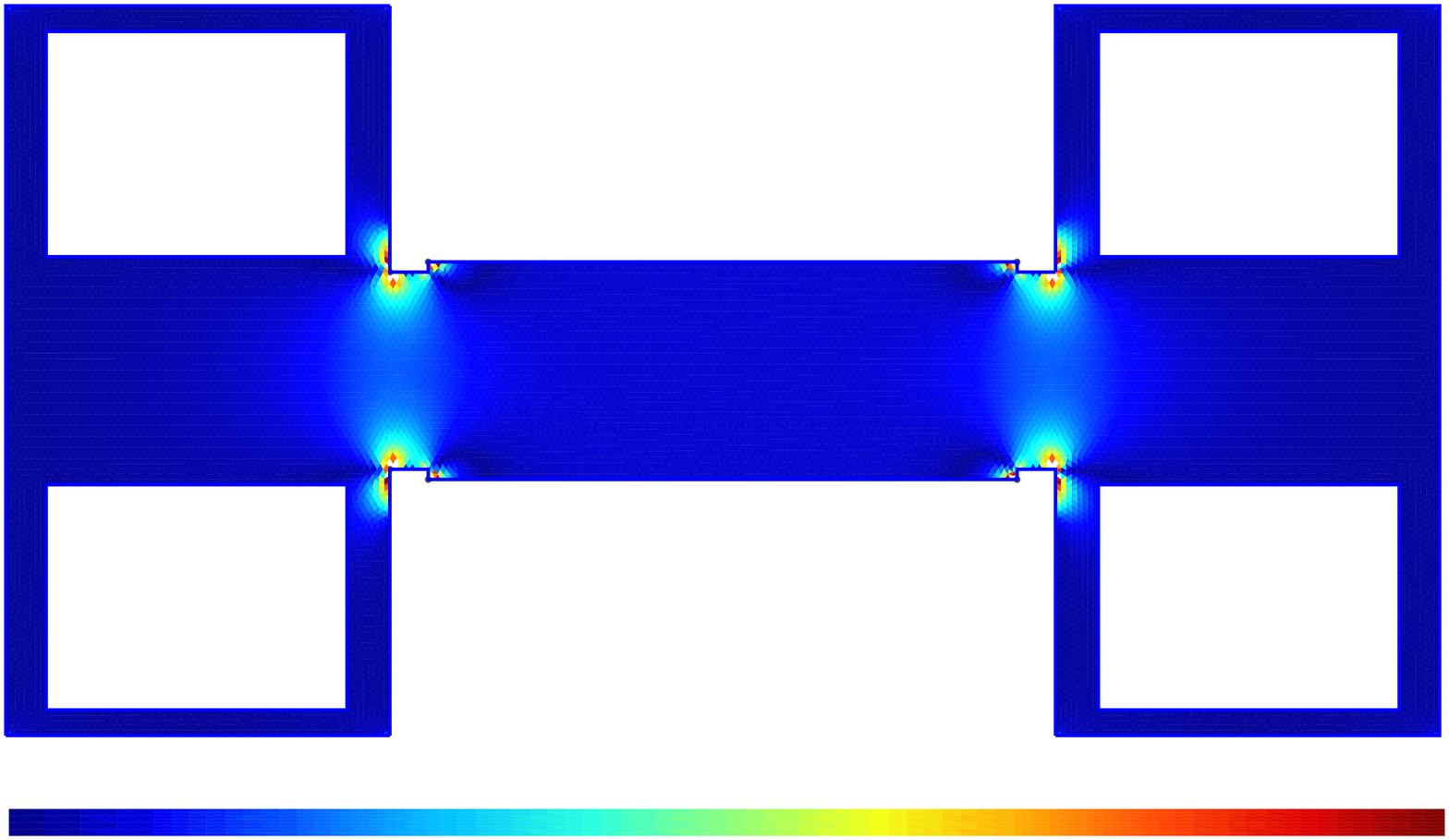\\
        \ \\
        \ \\
        \def\svgwidth{0.5\linewidth}
		\scriptsize
		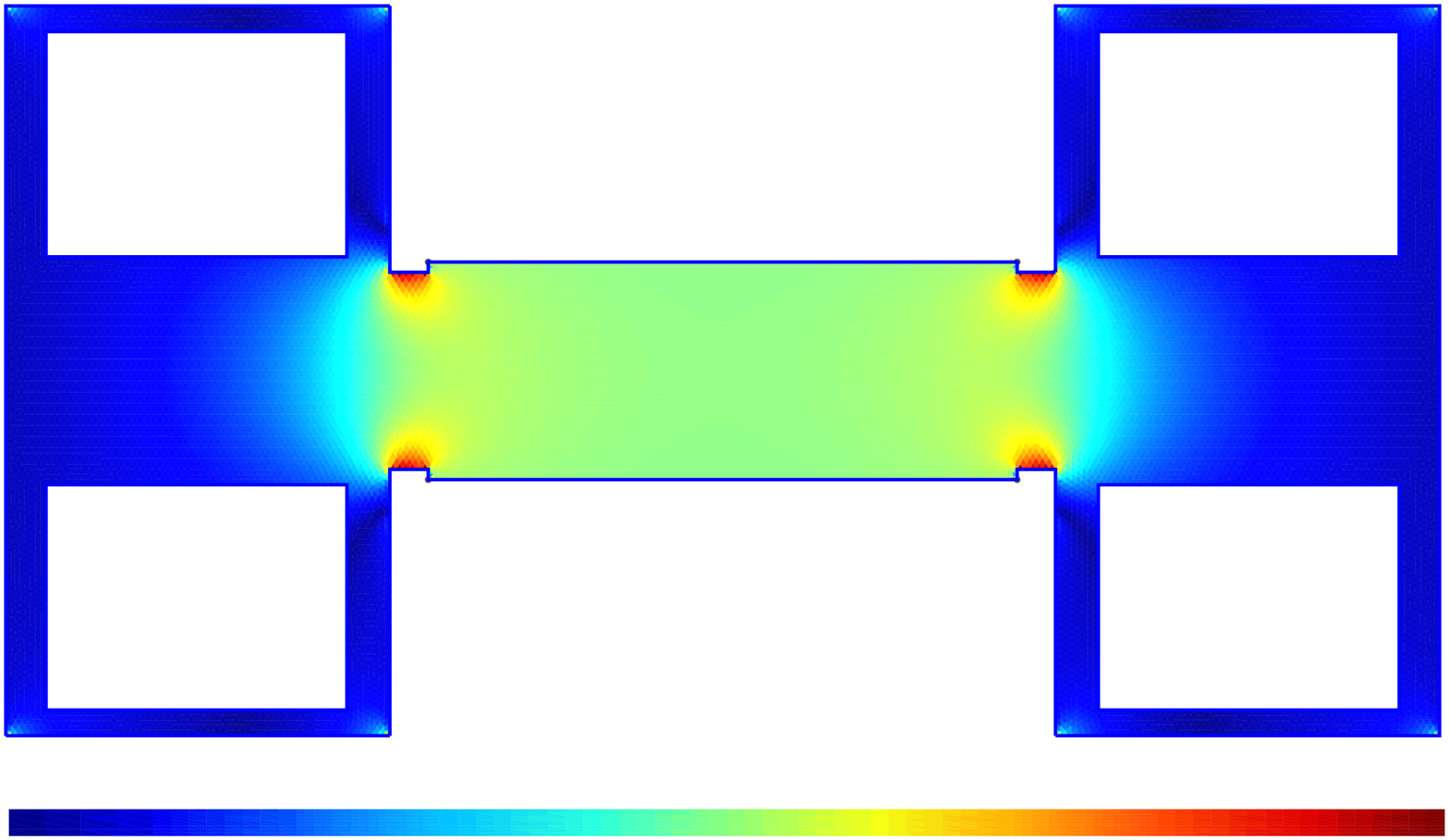  
			 \caption{Gateaux derivative of $\mathbf{B}_y$ with respect to perturbations of the reluctivity related to the weighted modes $\sqrt{\lambda_1} b_1$ and $\sqrt{\lambda_2} b_2$.\label{fig:sensitivity}}   
\end{figure}

For sensitivity analysis and parameter identification, the geometry of a H-shaped dipole (Fig.\ref{fig:yoke}) with 180 turns and a gap height of \SI{680}{\milli\m} is chosen as an example. A quantity of interest that is observable by measurements is the $\mathbf{B}_y$ component of the magnetic flux density. To determine measurement positions $\mathbf{x}_p$ where $\mathbf{B}_y$ is sensitive to changes of the reluctivity in $D_{\mathrm{iron}}$ related to the weighted modes $\sqrt{\lambda_m} b_m$ the perturbations $\tilde{\nu}_m(s):=\frac{1}{s}\sqrt{\lambda_m} b_m(s)$ are introduced and $\nu(s)=\frac{1}{s}\hat{f}_\mathrm{HB}(s)$ is considered as nominal reluctivity. The Gateaux derivative $\mathbf{B}_y'$ of the mapping
\begin{equation}
    \tilde{\nu}_m\mapsto\mathbf{B}_y\left[ \nu+\tilde{\nu}_m \right]
\end{equation}
increases towards the shims of the magnet (Fig.\ref{fig:sensitivity}). 
Therefore, magnetic flux density data $\mathbf{B}_n^\mathrm{data}(\mathbf{x}_p)$ in positions close to the shims is chosen as a reference in the objective function of the optimization problem. 

\begin{figure}
	\centering
	\input{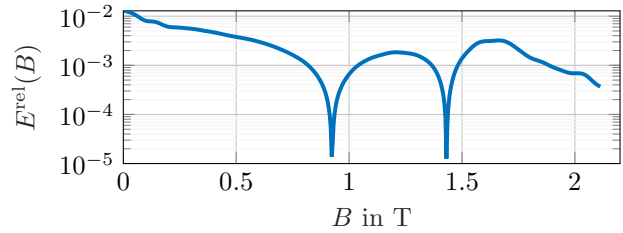}
	\caption{Relative error of $f_\mathrm{HB}(\arg\min\mathbf{y},\cdot)$ compared to the ground truth $f_\mathrm{HB}(\mathbf{y}_0,\cdot)$.}
	\label{fig:HB_relresult}
\end{figure}

\begin{figure}
	\centering
	\input{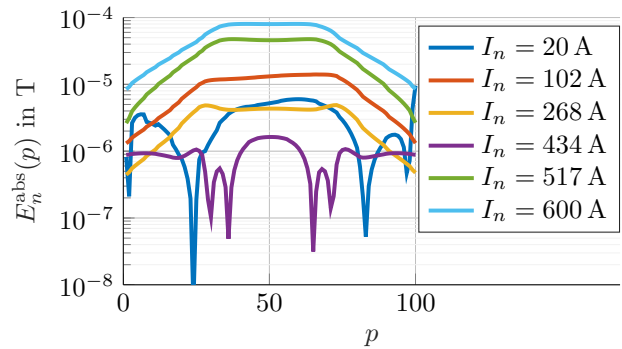}
	\caption{Absolute error $E^{\mathrm{abs}}_n(p)$ between the reconstructed magnetic flux density component $\mathbf{B}_{y,n}(\hat{\mathbf{x}}_p)$ based on $f_{\mathrm{HB}}(\arg\min\mathbf{y},\cdot)$ and the data $\mathbf{B}_{y,n}^\mathrm{data}(\hat{\mathbf{x}}_p)$ in reference positions $\hat{\mathbf{x}}_p$ along the central axis of the air gap.}
	\label{fig:By_result}
\end{figure}

To validate our parameter identification method, a ground truth parameter $\mathbf{y}_0\in[\mathbf{y}^\mathrm{min},\mathbf{y}^\mathrm{max}]$ is selected and simulated magnetic flux density data $\mathbf{B}_n^\mathrm{data}(\mathbf{x}_p)$ obtained by solving the corresponding magnetostatic problem in getDP for $n=1,\dots,8$ current levels $I_n\in[\SI{20}{\A},\SI{450}{\A}]$. The relative error
\begin{equation}
    E^{\mathrm{rel}}(B):=\frac{|f_{\mathrm{HB}}(\arg\min\mathbf{y},B)-f_{\mathrm{HB}}(\mathbf{y}_0,B)|}{f_{\mathrm{HB}}(\mathbf{y}_0,B)}
\end{equation}
of the resulting material curve $f_{\mathrm{HB}}(\arg\min\mathbf{y},\cdot)$ compared to the ground truth material curve $f_{\mathrm{HB}}(\mathbf{y}_0,\cdot)$ is less than 1.2\% (Fig.~\ref{fig:HB_relresult}). The absolute error
\begin{equation}
    E^{\mathrm{abs}}_n(p):=|\mathbf{B}_{y,n}(\hat{\mathbf{x}}_p)-\mathbf{B}_{y,n}^\mathrm{data}(\hat{\mathbf{x}}_p)|
\end{equation}
of the reconstructed magnetic flux density component $\mathbf{B}_{y,n}(\hat{\mathbf{x}}_p)$ based on \\ $f_{\mathrm{HB}}(\arg\min\mathbf{y},\cdot)$ compared to the data $\mathbf{B}_{y,n}^\mathrm{data}(\hat{\mathbf{x}}_p)$ in reference positions $\hat{\mathbf{x}}_p$ along the central axis of the air gap of the magnet remains below \SI{0.0001}{\tesla}, see Fig.~\ref{fig:By_result}. This result also holds for positions $\hat{\mathbf{x}}_p$ and current levels $I_n>\SI{450}{\A}$ that have not been part of the training data $\mathbf{B}_{n}^\mathrm{data}(\mathbf{x}_p)$ used in the objective function of the optimization problem.

%comparison with RGF relative values

\section{Conclusion}
In this paper, a data-driven inverse procedure to identify the parameters of a stochastic $B(H)$ curve model is described and applied to a test case. It is shown that a ground truth $B(H)$ curve can be retrieved with sufficient accuracy such that the resulting updated numerical model matches the data of the magnetic flux density in positions and current levels that were not part of the data set used for the update. Thus, the $B(H)$ curve of a numerical model can be tailored to the yoke of a magnet as built. Future research shall include measured instead of simulated data for the updating step. Then, also uncertainties in the data could be taken into account and therefore the inverse problem solved with the Bayesian approach instead of the optimization.

\section*{Acknowledgment}
This work has been supported by the Gentner Programme of the German Federal Ministry of Education and Research, and the Grad. School for Comp. Eng. at TU Darmstadt. The authors like to thank Ulrich Römer for the many fruitful discussions regarding the KLE and sensitivity analysis.

\end{document}